\documentclass[12pt]{article}

\usepackage{amsmath}
\usepackage{amssymb}
 
\newcommand{\CP}{\mathbb{P}}
\newcommand{\ns}{\ \indent}

\begin{document}

\title{\vskip -70pt
  \begin{flushright}
   {\normalsize JHEP 02(2000)042, {\tt hep-th/9910212}}
  \end{flushright}
  \vskip 15pt
{\bf On the generalized Legendre transform and monopole metrics}
\author{ {\large  C.J. Houghton},\\
{\normalsize {\sl Physics Department, Columbia University, New York,
    New York 10027, USA}}\\{\normalsize {\sl Email: houghton@cuphyb.phys.columbia.edu}}}
\date{{\large 14 December, 1999}}}
\maketitle

\begin{abstract}
In the generalized Legendre transform construction the K\"ahler
potential is related to a particular function. Here, the form of this
function appropriate to the $k$-monopole metric is calculated from
the known twistor theory of monopoles.
\end{abstract}

\section{Introduction}

\ns Monopole moduli spaces are hyperk\"ahler and, therefore, have a twistor
description.  Such a description is given in \cite{AH}. More recently,
Ivanov and Ro\v cek used the generalized Legendre transform of
\cite{LR} to construct the metric on the 2-monopole moduli space
\cite{IR}. The relationship between these two twistor
constructions is clarified in this paper.
 
The twistor space of a hyperk\"ahler manifold, $M$, is a trivial fiber
bundle, $Z=M\times\CP^1$, with a holomorphic symplectic form $\omega$,
which is an ${\cal O}(2)$ section over $\CP^1$. $\CP^1$ is covered by
two affine patches, $U_0$ and $U_1$. If $\zeta$ is the usual
projective coordinate on $U_0$, then $\omega$ is given on the fiber
over $\zeta$ by
$\omega=(\omega_2+i\omega_3)+2\zeta\omega_1-\zeta^2(\omega_2-i\omega_3)$
where $\omega_1$, $\omega_2$ and $\omega_3$ are the covariantly
constant 2-forms which hyperk\"ahlerity implies exist on $M$.

The generalized Legendre transform construction shows how the K\"ahler
potential may be calculated, if $\omega$ is assumed to have a certain
form. In subsection \ref{TgLt} of this introductory section, there is a
brief review of this construction. This is followed, in subsection
\ref{Tm}, by a brief review of the twistor theory of monopoles and the
related twistor theory of monopole moduli spaces.  In section
\ref{gLtmms}, this twistor theory is re-expressed as a generalized
Legendre transform. The Legendre transform constraints are then the
Ercolani-Sinha conditions \cite{ES,HMR}.

\subsection{Twistors and the generalized Legendre transform \label{TgLt}}

\ns The generalized Legendre transform construction of \cite{HKLR, LR,
IR} concerns twistor spaces with $k$ intermediate holomorphic
projections $Z\rightarrow {\cal O}(2n_j)\rightarrow \CP^1$, where
$j=1\ldots k$ and $n_j$ are integers. In the example of interest in
this paper $n_j=j$ and for ease of notation attention is restricted to
that case.  This requirement is equivalent to the existence of $k$
coordinates $\alpha^j(\zeta)$ on $Z$, so that $\alpha^j(\zeta)$ is a
degree $2j$ polynomial:
\begin{equation}
\alpha^j=\sum_{a=1}^{2j} w^j_a\zeta^a,
\end{equation}
satisfying the reality condition
$
\alpha^j(\zeta)=(-1)^j\zeta^{2j}\overline{\alpha^j(-1/\bar{\zeta})}.
$

The construction is derived from the patching formulae relating
quantities over $U_0$ and $U_1$. The coordinate on $U_1$ is given by
$\tilde{\zeta}=1/\zeta$. Since $\omega$ is an ${\cal O}(2)$ line
bundle, it is given on $U_1$ by $\tilde{\omega}$, where
\begin{equation}
\tilde{\omega}=\frac{1}{\zeta^2}\omega
\end{equation}
on $U_0\cap U_1$. Similarly, the $\alpha^j$ coordinates are related by
\begin{equation}
\tilde{\alpha}^j=\frac{1}{\zeta^{2j}}\alpha^j.
\end{equation}
This means that, if $(\alpha^j,\xi^j,\zeta)$ are coordinates for the
whole of $Z$, such that
\begin{equation}
\omega=\sum_{j=1}^k d\alpha^j\wedge d\xi^j,
\end{equation}
then, the patching formula for
$\xi^j$ must be
\begin{equation}\label{xipatch}
\tilde{\xi}^j=\zeta^{2j-2}\left(\xi^j+\frac{\partial H}{\partial \alpha^j}\right)
\end{equation}
for some function $H(\alpha^j)$.

The expansion of these coordinates as power series is now
considered. The patching formulae will place constraints on the values
of certain coefficients in these expansions and these constraints
will be unified as constraints on a single function $F$. By expanding the
symplectic form in $\zeta$, it is possible to calculate the K\"ahler
potential for the metric on $M$ in terms of $F$.

Assuming that $\xi^j$ is non-singular near $\zeta=0$; 
\begin{equation}
       \xi^j =\sum_{n=0}^{\infty}x^j_n\zeta^n,\qquad
\tilde{\xi}^j=\sum_{n=0}^{\infty}y^j_n\zeta^{-n}.
\end{equation}
Using the residue theorem and the patching formula (\ref{xipatch}),
this means that
\begin{equation}
x^j_m=\frac{1}{2\pi i}\oint_0\xi^j\frac{d\zeta}{\zeta^{m+1}}
     =\frac{1}{2\pi i}\oint_0\tilde{\xi}^j\frac{d\zeta}{\zeta^{m+2j-1}}-
         \frac{1}{2\pi i}\oint_0 \frac{\partial H}{\partial \alpha^j} \frac{d\zeta}{\zeta^{m+1}}
\end{equation} 
where $0$ is the small contour surrounding $\zeta=0$. The integral of
$\tilde{\xi}^j$ does not give the coefficient $y^j_{2-m-2j}$, because
the contour around $\zeta=0$ may enclose branch cuts. It is assumed
that the contribution from these cuts can be expressed as an integral
of some new function, $H'$, around some contour, $c$. This integral is
the effect of moving the contour of the $\tilde{\xi}^j$ integral from
a small contour around zero to a small contour around infinity. This
technique is justified by example and is dealt with carefully in the
specific case considered below. Thus,
\begin{equation}\label{compform}
x^j_m =y^j_{2-m-2j}- \frac{1}{2\pi i}\oint_c
\frac{\partial H'}{\partial \alpha^j} \frac{d\zeta}{\zeta^{m+1}}
-\frac{1}{2\pi i}\oint_0 \frac{\partial H}{\partial \alpha^j}
\frac{d\zeta}{\zeta^{m+1}}.
\end{equation}

A function $F$ is defined as
\begin{equation}
F= -\frac{1}{2\pi i}\oint_c H' \frac{d\zeta}{\zeta^2}
-\frac{1}{2\pi i}\oint_0 H \frac{d\zeta}{\zeta^2}.
\end{equation}
By the chain rule
\begin{equation}
\frac{\partial F}{\partial w^j_n} =-\frac{1}{2\pi i}\oint_c
\frac{\partial H'}{\partial \alpha^j} \zeta^n\frac{d\zeta}{\zeta^2}
-\frac{1}{2\pi i}\oint_0 \frac{\partial H }{\partial \alpha^j}
\zeta^n\frac{d\zeta}{\zeta^2}
\end{equation}
Therefore, from (\ref{compform})
\begin{equation}\label{aj1}
\frac{\partial F}{\partial w^j_0}=x_1^j,\qquad
 \frac{\partial
F}{\partial w^j_1}=x_0^j
\end{equation}
and, for $0 < a < 2j-2$,
\begin{equation}\label{aja}
\frac{\partial F}{\partial w^j_a}=0.
\end{equation}

The symplectic form can be expanded in $\zeta$
to determine $\omega_1$ on the fiber above $\zeta=0$. The coordinates on this fiber are $\alpha^j(0)=w^j_0$ and $\xi^j(0)=x^j_0$. It can then be shown that
the K\"ahler form for the hyperk\"ahler manifold, $M$, is given by
\begin{equation}
K(w^j_0,x^j_0)=F(w^j_0,w^j_1)-\sum_{j=1}^k x^j_0w^j_1-\sum_{j=1}^k \overline{x}^j_0\overline{w}^j_1
\end{equation}
where $w^j_1$ is related to $x^j_0$ by (\ref{aj1}) \cite{HKLR,IR}.
This is the generalized Legendre transform construction.

\subsection{Twistors and monopoles \label{Tm}}

\ns The twistor theory of monopoles is described in
\cite{Hi1,Hi2}. A $k$-monopole is equivalent to a curve, $S$, in $T\CP^1$ of
the form
\begin{equation}\label{sc}
P(\eta,\zeta)=\eta^k+\sum_{j=1}^{k}\alpha^j\zeta^{k-j}=0
\end{equation}
where $\alpha^j$ is, as before, a degree $2j$ polynomial satisfying
the reality condition. $\eta$ is the usual coordinate on the fiber of
$T\CP^1\rightarrow \CP^1$. $S$ is called the spectral curve. In
addition to the reality condition on $\alpha^j$, it must also satisfy
a number of algebraic-geometric conditions. In particular it is
required that the $L^2$ bundle over the spectral curve must be
trivial.

The patches $U_0$ and $U_1$ on $\CP^1$ lift to patches on $T\CP^1$.
These, in turn, give patches on the spectral curve: these will also be
called $U_0$ and $U_1$. The triviality of the $L^2$ bundle over the
spectral curve, means that there is a section given by two nowhere
vanishing holomorphic functions $f_0$ on $U_0$ and $f_1$ on $U_1$
satisfying
\begin{equation}\label{L2}
f_0(\eta,\zeta)=e^{-\frac{2\eta}{\zeta}}f_1(\eta,\zeta)
\end{equation}
on the intersection $U_0\cap U_1$.

In \cite{ES}, explicit integral conditions are given for the existence
of such a section. There must exist a 1-cycle $c$ so that 
any global holomorphic 1-form, $\Omega$, satisfies
\begin{equation}\label{L2triv}
\oint_c\Omega=-2\sum_{j=1}^{k}\eta_j(0)g_j
\end{equation}
where $\eta_j(0)$ is the value of $\eta$ at $0_j$, the point on the
$j$th sheet above $\zeta=0$. $g_j$ is defined by writing $\Omega=g_j
d\zeta$ at $0_j$. These are the Ercolani-Sinha conditions. $c$
must be primitive if the $L^s$ bundle on $S$ is nontrivial for
$0<s<2$: another necessary condition.

In the calculation of section \ref{gLtmms}, the relationship between
$c$ and the $L^2$ section will be important. If
$\{a_1,\ldots,a_g,b_1,\ldots,b_g\}$ is a canonical homology basis,
\begin{equation}\label{c}
c=\sum_{r=1}^{g}(n_ra_r+m_rb_r)
\end{equation}
where $n_r=\frac{1}{2\pi i}\oint_{b_r}d\log{f_1}$ and
$m_r=-\frac{1}{2\pi i}\oint_{a_r}d\log{f_1}$ are integers.

The twistor data for monopoles can also be used to derive a twistor
space for the $k$-monopole moduli space, $M_k$.  Coordinates for $M_k$
are provided by the rational map description \cite{D}. This
description relates a $k$-monopole to a degree $k$ based rational map:
$p(z)/q(z)$. $q(z)$ is a monic polynomial of degree $k$ and
$p(z)$ is a polynomial of degree $k-1$, which has no factors in common
with $q(z)$.  Following Hurtubise \cite{Hu2}, the rational map for a
monopole can be constructed from the spectral curve and the
trivialization of the $L^2$ bundle. A direction, $\zeta$, is chosen
and
\begin{eqnarray}
q(z;\zeta)&=&P(z,\zeta)\nonumber\\
p(z;\zeta)&\equiv& f_0(z,\zeta)\quad \mbox{mod}\;q(z).
\end{eqnarray}

Now, if $q(z)$ has distinct roots, $\eta_1,\ldots,\eta_k$, coordinates
for $M_k$ are given by
$(\eta_1,\ldots,\eta_k,p(\eta_1),\ldots,p(\eta_k))$. Atiyah and
Hitchin point out in \cite{AH}, that the symplectic form
\begin{equation}
\omega=\sum_{i=1}^k\frac{dp(\eta_i)\wedge d\eta_i}{p(\eta_i)}
\end{equation}
has the property that $\tilde{\omega}=\zeta^{-2}\omega$ and is, therefore, an
${\cal O}(2)$ section over $\CP^1$.

In the next section, the relationship between the rational map and the
spectral curve is exploited to clarify the relationship between this
symplectic form and the generalized Legendre transformation.

\section{The generalized Legendre transform and monopole moduli spaces \label{gLtmms}}

\ns The formula for the spectral curve (\ref{sc}) defines $\eta_i$ as
roots of a polynomial equation whose coefficients are ${\cal O}(2j)$
sections. Because of this, $\omega$ can be rewritten
\begin{equation}
\omega=\sum_{i=1}^k \sum_{j=1}^{k-1} \frac{dp(\eta_i)}{p(\eta_i)}\wedge
\frac{\partial \eta_j}{\partial
\alpha^j}d\alpha^j=\sum_{j=1}^{k-1}d\xi^j\wedge d\alpha^j
\end{equation}
where
\begin{equation}\label{xitoalpha}
\xi^j=\sum_{i=1}^k\frac{\partial \eta_i}{\partial \alpha^j} \chi(\eta_i)
\end{equation}
and $\chi(\eta,\zeta)=\log{p}(\eta;\zeta)$. $\xi^j$ is not defined on the spectral curve, since $\chi$ is not. In general, $\oint_a d\chi\not=0$ for a nontrivial cycle $a$. $\chi$ can be defined by cutting the surface. The 1-forms $d\chi$ and $d\xi^j$ are defined on the uncut surface.

The patching formula for $\chi$ follows from the $L^2$ patching
formula (\ref{L2}), since $\tilde{\chi}=\log{f_1}$,
\begin{equation}
\tilde{\chi}=\chi+\frac{2\eta}{\zeta}.
\end{equation}
This, in turn, provides a patching formula for $\xi^j$,
\begin{equation}
\tilde{\xi}^j=\zeta^{2j-2}\left(\xi^j+2\sum_{i=1}^{k}\frac{\partial \eta_i}{\partial \alpha^j}\frac{\eta_i}{\zeta}\right)=\zeta^{2j-2}\left(\xi^j+\frac{\partial}{\partial \alpha^j}\sum_{i=1}^{k}\frac{\eta_i^2}{\zeta}\right).
\end{equation}
Therefore, $\xi^j$ has a Legendre transform patching formula with
\begin{equation}
H=\sum_{i=1}^k\frac{\eta_i^2}{\zeta}.
\end{equation}
This is the Hamiltonian function mentioned in \cite{AH}.

Now, as before, the integral around $\zeta=0$ is considered. Because of the 
particular form of the sum in the expression for $H$, the integral
on $\CP^1$ can be written as an integral on the spectral curve.
\begin{equation}
\frac{1}{2\pi i}\oint_0 \frac{\partial }{\partial \alpha^j} H
\frac{d\zeta}{\zeta^{m+1}}=\frac{1}{2\pi i}\oint_0 \frac{\partial }{\partial \alpha^j} \sum_{i=1}^k\frac{\eta_i^2}{\zeta}  \frac{d\zeta}{\zeta^{m+1}}
                          = \frac{1}{2\pi i}\oint_{\sum_{i=1}^k 0_i}\frac{\partial }{\partial \alpha^j}\frac{\eta^2}{\zeta} \frac{d\zeta}{\zeta^{m+1}}
\end{equation}
where $0_j$ is the small contour on the $j$th sheet of the spectral curve around the lift of $\zeta=0$ to that sheet.

Next, the contour in the $\tilde{\xi}^j$ integral must be moved from $0$ to
$\infty$. This is not difficult if the integral is first rewritten as an
integral on the spectral curve:
\begin{equation}
\frac{1}{2\pi i}\oint_0 \tilde{\xi}^j\frac{d\zeta}{\zeta^{m+2j-1}}=
\frac{1}{2\pi i}\oint_{\sum_{i=1}^k 0_i} \frac{\partial
\tilde{\eta}}{\partial \tilde{\alpha}^j}\tilde{\chi}\frac{d\zeta}{\zeta^{m+2j-1}}
\end{equation}
$\tilde{\chi}$ is defined on the $4g$-gon, formed by cutting the
spectral curve along the canonical homology 1-cycles $a_r$ and $b_r$
for $r=1\dots g$. Although the spectral curve is obtained from the
$4g$-gon by identifying appropriate edges, the function $\tilde{\chi}$
does not respect the identifications. In fact, since, for example,
\begin{equation}
\oint_{b_1} d\log{f_1}=2\pi i n_1
\end{equation}
the value of $\tilde{\chi}$, at a point on the $a_1^{-1}$ edge, is $2\pi n_1$ larger  than its value at the corresponding point on the $a_1$
edge. This means that
\begin{equation}
\frac{1}{2\pi i}\oint_{\mbox{{\scriptsize edge}}}f(\zeta)\tilde{\chi}d\zeta
   =-\sum_{r=1}^{g}\left(n_r\oint_{a_r} f(\zeta)d\zeta
                       +m_r\oint_{b_r} f(\zeta)d\zeta\right)
   =-\oint_c f(\zeta)d\zeta
\end{equation}
where $f(\zeta)$ is any function which is well-behaved on the edge and
$c$ is the special homology cycle mentioned above
(\ref{c}). Furthermore, it is easy to see that
\begin{equation}
\sum_{i=1}^{k}0_i+\sum_{i=1}^{k}\infty_i=\mbox{edge}
\end{equation}
and so
\begin{eqnarray}
\frac{1}{2\pi i}\oint_0 \tilde{\xi}^j\frac{d\zeta}{\zeta^{m+2j-1}}&=&
\frac{1}{2\pi i}\oint_\infty \tilde{\xi}^j\tilde{\zeta}^{m+2j-3}d\zeta
+\oint_c\frac{\partial\tilde{\eta}}{\partial
\tilde{\alpha}^j}\frac{d\zeta}{\zeta^{m+2j-1}}\nonumber\\
&=&\frac{1}{2\pi i}\oint_\infty
\tilde{\xi}^j\tilde{\zeta}^{m+2j-3}d\zeta
+\oint_c\frac{\partial\eta}{\partial\alpha^j}\frac{d\zeta}{\zeta^{m+1}}.
\end{eqnarray}

Thus, the Legendre transformation of the $k$-monopole metric is given
by
\begin{equation}\label{F}
F=-\oint_c\frac{\eta}{\zeta^2}d\zeta-\frac{1}{2\pi
i}\oint_{\sum_{i=1}^k 0_i}\frac{\eta^2}{\zeta^3} d\zeta.
\end{equation}
This $F$ has also been considered by Roger Bielawski \cite{B1}.

$F$ is composed of integrals on the spectral curve, rather that on
$\CP^1$ itself. The integrals can be rewritten as integrals on
$\CP^1$, although they become less succinct. If the integral in the
$k=2$ case is rewritten in this way the $F$ used by Ivanov and Ro\v
cek to calculate the Atiyah-Hitchin metric \cite{IR} is recovered. In
that case the two branches over $\zeta=0$ differ only by a sign in
$\eta$ and the special contour, $c$, is an equator.

In the $k=2$ case, the constraint (\ref{aja}) arising in the generalized
Legendre transform construction is precisely the one that Hurtubise
demonstrates must be satisfied for a spectral curve to ensure
triviality of the $L^2$ bundle \cite{Hu1}. In fact, it is true for all
$k$, that the generalized Legendre transformation constraints are the
$L^2$ triviality conditions (\ref{L2triv}).
  
This is demonstrated by using the spectral curve equation to rewrite the
integrands in the constraint equations.
\begin{equation}
\frac{d}{dw^j_a}P(\eta,\zeta)=\frac{\partial P}{\partial \eta} \frac{\partial \eta}{\partial w^j_a}+\frac{\partial P}{\partial w^j_a}=0
\end{equation}
implies that
\begin{equation}
\frac{\partial \eta}{\partial w^j_a}=-\frac{\zeta^{a}\eta^{k-j}}{\partial P/\partial\eta}.
\end{equation}
Therefore, the constraint equation requires that
\begin{equation}
-\frac{1}{2\pi i}\oint_{\sum_{i=1}^k 0_i}\frac{2\eta^{k-j+1}\zeta^{a-2}}{\partial P/\partial\eta}\frac{d\zeta}{\zeta}
=
\oint_c\frac{\eta^{k-j}\zeta^{a-2} d\zeta}{\partial P/\partial\eta}
\end{equation}
where $2\le a\le 2j-2$, or, put another way
\begin{equation}
\oint_c \Omega^{ja}=-\frac{1}{2\pi i}\sum_i\oint_{0_i}\frac{2\eta\Omega^{ja}}{\zeta}
\end{equation}
where
\begin{equation}
\Omega^{ja}=\frac{\eta^{k-j} \zeta^{a-2} d\zeta}{\partial P/\partial \eta}
\end{equation}
is a global holomorphic 1-form. In fact, these 1-forms form a basis
for the global holomorphic 1-forms on the spectral curve.  There,
using the residue theorem on the right-hand side of the equations shows
that it is the Ercolani-Sinha constraint (\ref{L2triv}). It may be
noted that the right hand side of this equation is actually zero
for $j\not=1$ \cite{HMR}.

Thus, the generalized Legendre transform for monopole moduli space can
be derived from the twistor description of monopoles.  The constraints
on $F$ are the Ercolani-Sinha constraints. However, it should be
emphasized that the Ercolani-Sinha constraints only ensure that the
$L^2$ bundle is trivial and that the $L^s$ bundle is not trivial if
$s<2$. These are necessary conditions.  They are not sufficient. The
additional condition requires that $H^0(S,L^s(k-2))=0$ for $0<s<2$. It
is possible that this condition may also be interpreted in terms of
the generalized Legendre transformation.

The crucial step in calculating the constraints on $F$, is the derivation
of the coefficient form of the patching equation by expanding the
patching formula and moving the contour in the $\tilde{\xi}^j$
integral. Mimicking this derivation also clarifies the relationship,
explained in \cite{HMR}, between the Ercolani-Sinha and
Corrigan-Goddard conditions \cite{CG}.

\subsection{The point dyon limit}

\ns In this paper $F$ has been calculated from the known $k$-monopole
symplectic form. In \cite{IR} Ivanov and Ro\v cek found $F$ for the
2-monopole metric by making a guess based on the known asymptotic
metric and then verifying this guess by explicit calculation of the
K\"ahler form. In fact, the asymptotic metric has been calculated for
$k$ monopoles by examining the dynamics of point dyons \cite{GM}, this
approximation was confirmed in \cite{B}. It is a simple matter to
derive this asymptotic metric from $F$, thereby reversing, for general
$k$, the original derivation of $F$ for $k=2$.

The spectral curve for a single monopole located at
$(\mbox{Re}z,\mbox{Im}z,x)$ is
\begin{equation}
\eta-z+2x\zeta+\bar{z}\zeta^2=0.
\end{equation}
This is the sphere in $T\CP^1$ corresponding to all the lines through
the monopole location.  The spectral curve for $k$-monopoles located
at well separated points $(\mbox{Re}z_i,\mbox{Im}z_i,x_i)$ is
approximated with exponential accuracy by the product of spheres \cite{B}
\begin{equation}
\prod_{i=1}^k(\eta-\gamma_i)=0
\end{equation}
where
\begin{equation}
\gamma_i=z_i-2x_i\zeta-\bar{z}_i\zeta^2.
\end{equation}
The $i$ and $j$ spheres touch at two points, the two roots of
$\gamma_i=\gamma_j$:
\begin{equation}
\zeta_{ij}^{\pm}=\frac{x_{ij}\pm\sqrt{x_{ij}^2+|z_{ij}|^2}}{\bar{z}_{ij}}
\end{equation}
where $z_{ij}=z_i-z_j$ and $x_{ij}=x_i-x_j$.  It is known that the
special contour $c$ must change sign under the reality transformation
$\zeta\rightarrow-1/\bar{\zeta}$ \cite{HMR} and so must be a sum of
contours which run from $\zeta_{ij}^-$ to $\zeta_{ij}^+=\zeta_{ji}^-$
on sphere $i$ and then from $\zeta_{ji}^-$ back to
$\zeta_{ji}^+=\zeta_{ij}^-$ on sphere $j$. In order for $F$ to
generate the asymptotic metric $c$ must be a sum of all such contours.
 
$F$ can be rewritten in terms of integrals on $\CP^1$. It is
\begin{equation}
F=-\sum_{i\not=j}\int_{ij}\frac{\gamma_{ij}}{\zeta^2}d\zeta-\sum_i \frac{1}{2\pi i}\oint_0\frac{\gamma_i^2}{\zeta^3}d\zeta
\end{equation}
where $\gamma_{ij}=\gamma_i-\gamma_j$ and the $ij$ integral is along
the line running from $\zeta_{ij}^-$ to $\zeta_{ij}^+$. In order to
change the line integrals into contour integrals a $\log{\gamma_{ij}}$
is introduced into the integrand, thus,
\begin{equation}
F=-\sum_{i\not=j}\frac{1}{2\pi i}\oint_{ij}\frac{\gamma_{ij}\log{\gamma_{ij}}}{\zeta^2}d\zeta
-\sum_i\frac{1}{2\pi i}\oint_0\frac{\gamma_i^2}{\zeta^3}d\zeta
\end{equation}
where the $ij$ integral is now around the figure of eight contour
enclosing the two zeros of $\gamma_{ij}$. This $F$ has been discussed
in \cite{C} and gives the correct asymptotic metric.

\section{Conclusions}

The function $F$ appropriate to the generalized Legendre transform
construction of multimonopole metrics is calculated by a simple change
of variables. This function is a contour integral over the spectral
curve.  The constraints on $F$ are precisely the integral constraints
on the spectral curve required to ensure the existence of a trivial
$L^2$ bundle. In practice these constraints are difficult to apply.

The generalized Legendre transform was originally derived from a
duality transformation on an N=4 supersymmetric $\sigma$-model. It
would be interesting to understand what relationship this
$\sigma$-model has to monopoles.

\section*{Acknowledgments} 
\ns I have had many benificial conversations with N.S. Manton and
N.M. Rom\~ao regarding spectral curves. I also thank N.J. Hitchin and
P.M. Sutcliffe for useful remarks. I am grateful to the Fulbright
Commission and the Royal Commission for the Exhibition of 1851 for
financial support.

\end{document}